\RequirePackage{lineno}

\documentclass[twocolumn,showpacs,aps,prl,superscriptaddress]{revtex4}

\usepackage{xspace}
\usepackage{graphicx} 
\usepackage{dcolumn}  
\usepackage{colordvi}
\usepackage{color}
\usepackage{epstopdf}
\usepackage{amsmath}
\usepackage{epsfig}
\usepackage{multirow}
\usepackage{amssymb}
\usepackage{url}
\graphicspath{{ps}}
\usepackage{hyperref}
\usepackage{verbatim}
\usepackage{pgffor}
\usepackage{orcidlink}
\usepackage{ulem}

\input{belle2sym.tex}

\begin{document}
\def\belletwo {\it {Belle II}}

\vspace*{-3\baselineskip}
\resizebox{!}{3cm}{\includegraphics{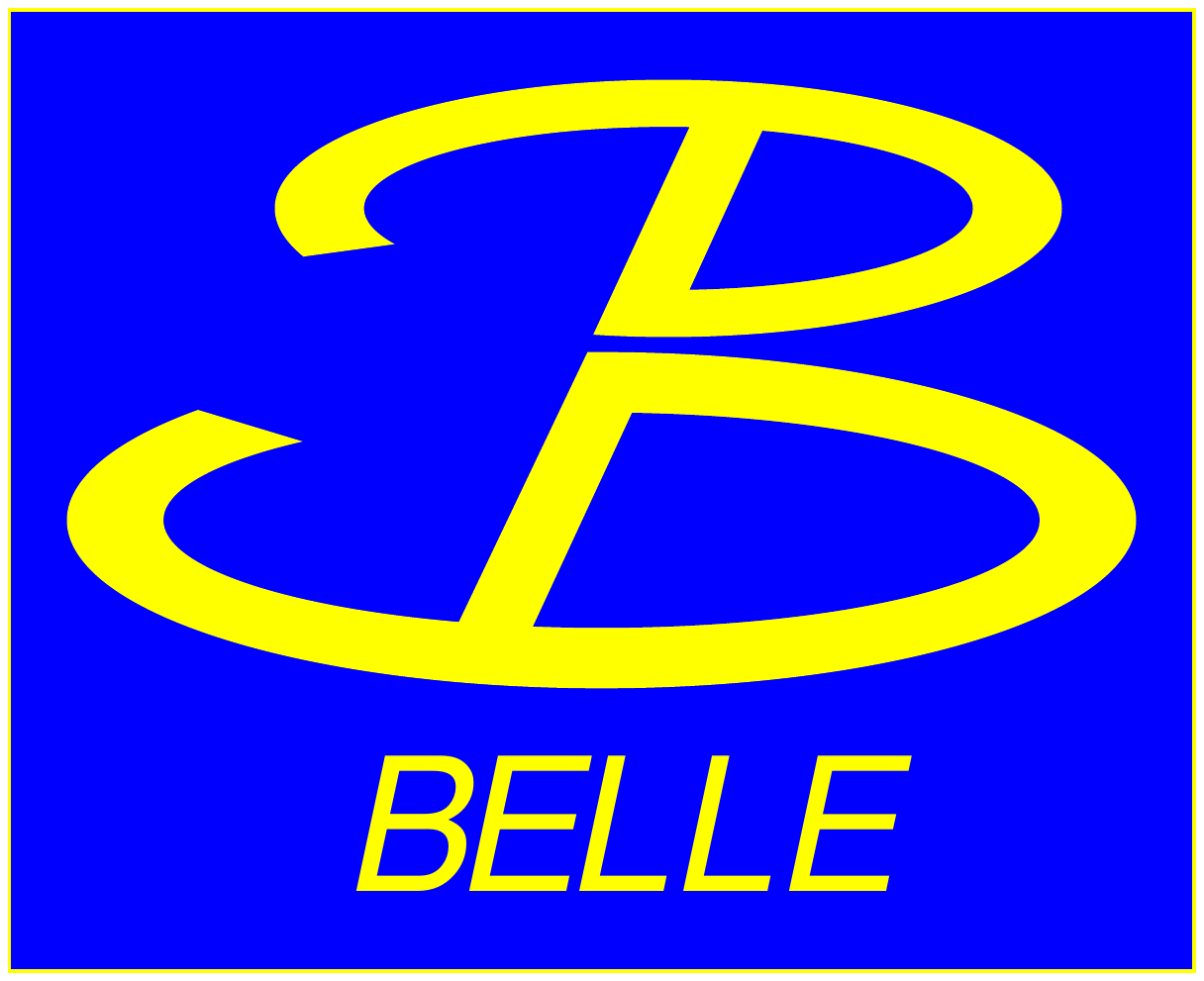}}

\vspace*{-6\baselineskip}
\begin{flushright}
Belle Preprint 2023-13\\
KEK  Preprint 2023-17\\
\today
\end{flushright}

\vspace*{-3\baselineskip}

\title { \quad\\[0.5cm] Search for a dark leptophilic scalar produced in association with a \tautau pair in \epem annihilation at center-of-mass energies near 10.58~\gev}

\noaffiliation
  \author{D.~Biswas\,\orcidlink{0000-0002-7543-3471}} 
  \author{Sw.~Banerjee\,\orcidlink{0000-0001-8852-2409}} 
  \author{I.~Adachi\,\orcidlink{0000-0003-2287-0173}} 
  \author{H.~Aihara\,\orcidlink{0000-0002-1907-5964}} 
  \author{D.~M.~Asner\,\orcidlink{0000-0002-1586-5790}} 
  \author{T.~Aushev\,\orcidlink{0000-0002-6347-7055}} 
  \author{R.~Ayad\,\orcidlink{0000-0003-3466-9290}} 
  \author{V.~Babu\,\orcidlink{0000-0003-0419-6912}} 
  \author{P.~Behera\,\orcidlink{0000-0002-1527-2266}} 
  \author{J.~Bennett\,\orcidlink{0000-0002-5440-2668}} 
% ...
\collaboration{The Belle Collaboration}

\begin{abstract}
A dark leptophilic scalar $(\phi_L)$ is a hypothetical particle
that couples only to leptons rather than quarks.
We report on a search for 
$\phi_L$ in the $e^+e^- \to \tau^+ \tau^- \phi_L, ~\phi_L \to \ell^+ \ell^- ~(\ell = e, \mu)$ 
process using 626~\invfb of data collected by the Belle experiment 
near the \Y4S resonance. We validate the backgrounds 
with multiple control regions in data, 
using a novel multiclass multivariate event classifier. 
In absence of a signal, 
we quote upper limits at the 90\% confidence level
on the coupling between $\phi_L$ and leptons.
Our bounds, obtained in a blinded approach, 
are 19\% more constraining than the previous limits,
averaged over the mass range $0.04 \leq m_{\phi_L} \leq 6.5~\gev$.
We exclude the parameter space below 4~\gev favored by measurement of the anomalous magnetic moment of the muon.

\end{abstract}


\maketitle



The astrophysical observation of the dark matter in the universe~\cite{Corbelli:1999af},
and measured excess over Standard Model (SM) expectations
in the anomalous magnetic moment of the muon, $(g-2)_\mu$~\cite{Muong-2:2021ojo}, 
could be signatures of new physics beyond the SM.
Recently, models with a dark leptophilic scalar ($\phi_L$), 
which couples directly only to leptons~\cite{Fox:2008kb, Chen:2015vqy},
have been introduced at mass scales substantially lighter 
than the weak scale.
Models, in which a generic dark scalar ($\phi)$ couples to quarks
as well, are strongly constrained by the existing limits on
the decays through flavor-changing neutral current, such as,
$B \to K \phi$ and $K \to \pi \phi $~\cite{Abercrombie:2015wmb, Beacham:2019nyx}.
The leptophilic models evade most of such existing bounds
with a minimal scenario that includes a mixing 
between $\phi_L$ and the SM particles ~\cite{Branco:2011iw, Liu:2016qwd}.
These models can explain the observed excess 
in measured $(g-2)_\mu$~\cite{Agrawal:2014ufa, 
Batell:2016ove, Liu:2020qgx},
violation of lepton flavor universality~\cite{Calibbi:2015sfa, Crivellin:2020klg}, 
or recent hints of new physics in a model-independent framework~\cite{Freitas:2014jla}.

In this model, mixing between $\phi_L$ and the Higgs boson 
produces couplings proportional to fermion masses,
described by the following term in the Lagrangian
~\cite{Batell:2016ove}:
\begin{equation}
{\mathcal{L}}=-\xi \sum_{\ell=e,\mu,\tau} \frac{m_{\ell}}{v}\bar{\ell}\phi_L\ell,
\end{equation}
where $\xi$ denotes the strength of flavor-independent coupling 
to leptons ($\ell$) with mass $m_{\ell}$, and $v=246~\gev$~\cite{NaturalUnits} 
is the vacuum expectation value of the Higgs field.

\begin{figure}[!htbp]
\includegraphics[width=.47\textwidth]{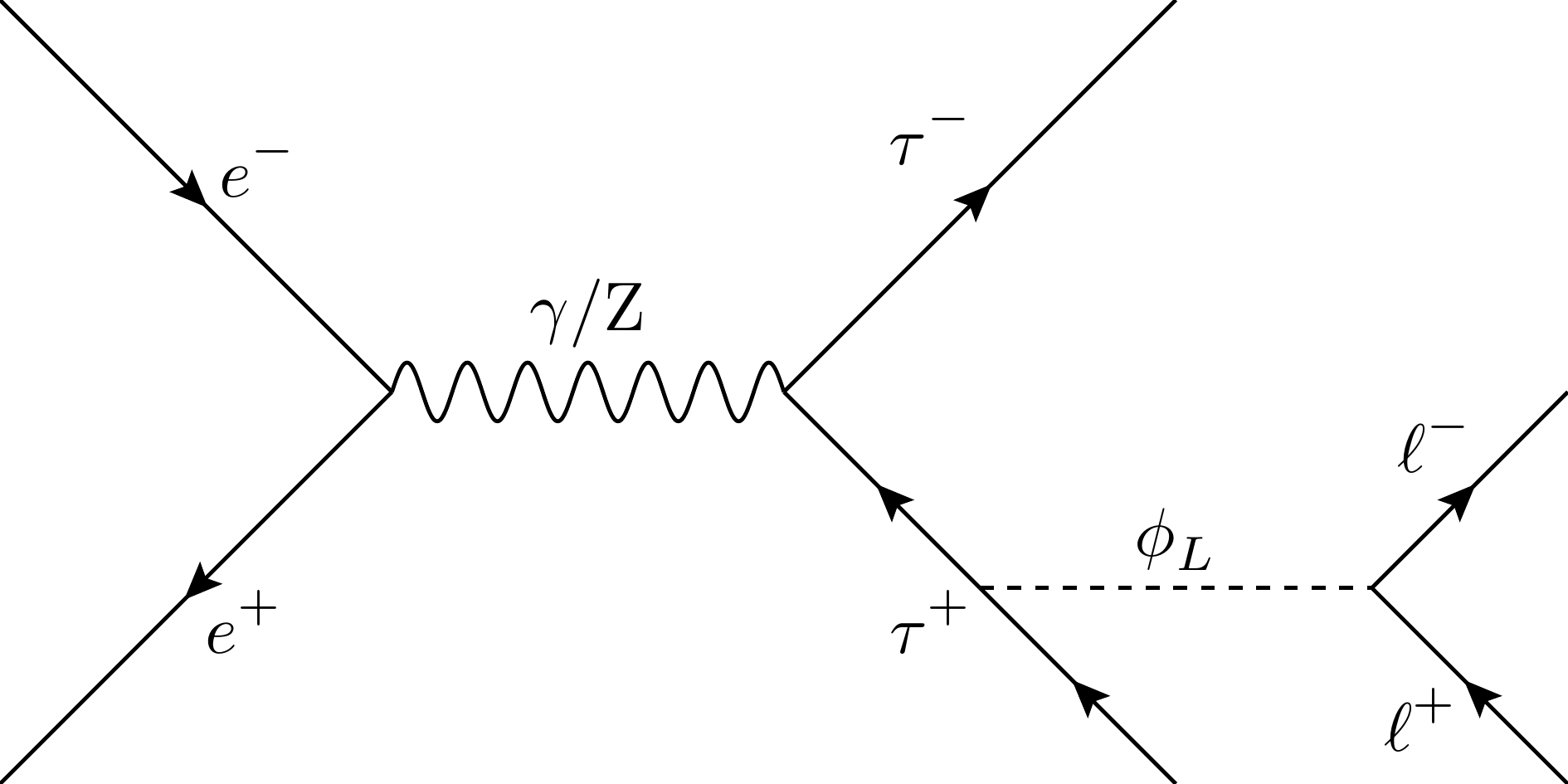}
\caption{Dominant Feynman Diagram for production of $\phi_L$
in association with \tautau pair in \epem annihilation.
}
\label{fig_FeynmanDiagram}
\end{figure}

Here, we report a search for a leptophilic scalar in the process 
$e^+e^-\to \tau^+\tau^-\phi_L, ~\phi_L\to \ell^+\ell^- ~(\ell=e,\mu)$.
The dominant Feynman diagram is shown in Fig.~\ref{fig_FeynmanDiagram}.

The cross-section of 
$e^+e^-\to \tau^+\tau^-\phi_L, ~\phi_L \to \epem$ 
falls sharply beyond the di-muon
threshold,
where the $\phi_L \to \mumu$ channel opens up~\cite{Batell:2016ove}. 
We search in $\phi_L \to e^+e^-$ channel 
only up to $\phi_L$ mass $m_{\phi_L} = 2m_\mu$,
and $\phi_L \to \mumu$ channel for $m_{\phi_L} > 2m_\mu$.
Although for $m_{\phi_L} > 2m_\tau$, the cross-section
of the $e^+e^-\to \tau^+\tau^-\phi_L, ~\phi_L \to \mumu$ process 
decreases~\cite{Batell:2016ove}, 
we are still able to set competitive limits till $m_{\phi_L} = 6.5~\gev$.


The data used in this analysis was recorded by the Belle experiment
from the collision of 8~\gev electrons with 3.5~\gev positrons
at the KEKB collider~\cite{Akai:2001pf}.
The Belle detector, a large-solid-angle magnetic spectrometer, 
is described in detail elsewhere~\cite{Belle:2000cnh}.
Outward from 15~\mm radius beam pipe~\cite{Abe:2004ma},
it consists of a four-layer silicon vertex detector (SVD),
a 50-layer central drift chamber (CDC),
an array of aerogel threshold Cherenkov counters (ACC),
a barrel-like arrangement of time-of-flight scintillation counters,
and an electromagnetic calorimeter comprised of CsI(Tl) crystals (ECL),
all located inside a superconducting solenoid coil 
that provides a 1.5~T magnetic field.
Clean electron identification is obtained 
by combining the responses of the ECL, CDC, and ACC detectors,
while muons are identified by CDC and resistive plate chambers
in the instrumented iron flux-return 
located outside the coil.


The data-set corresponds to a luminosity of 626~\invfb
collected after the upgrade of the SVD sub-detector in October 2003.
Out of these, 562~\invfb was collected at the \Y4S resonance
and the remaining at a center-of-mass (c.m.) energy 60~\mev 
below the resonance.
The luminosity values are measured 
with a relative systematic uncertainty 
of 1.4\%~\cite{Belle:2012iwr}.
The  $\epem\to\qqbar$
(where $q = u,~ d,~ s~\mathrm{~or~} c$),
and $\epem\to\BB$ Monte Carlo (MC) samples
are generated with {\tt EvtGen}~\cite{Lange:2001uf}.
The $\epem\to\epem$ and
$\epem\to\epem(\ellell/\qqbar)$ (two-photon) 
samples are generated using
{\tt BHLUMI}~\cite{Jadach:1991by} and
{\tt AAFHB}~\cite{Berends:1986ig}, respectively.
We use {\tt KKMC}~\cite{Jadach:1999vf} to generate 
$\epem\to\mumu$ and $\epem\to\tautau$ processes, 
and {\tt TAUOLA}~\cite{Jadach:1993hs} 
to subsequently decay the $\tau$ leptons. 
Final state radiation is modeled with {\tt PHOTOS}~\cite{Barberio:1993qi}.
The signal process, $\epem\to\tautau\phi_L,~\phi_L\to\ellell$,
is generated by a new feature 
of {\tt PHOTOS++}~\cite{Banerjee:2021rtn} integrated into {\tt KKMC}.
The signal cross-sections are calculated 
using {\textsc{MadGraph~5}}~\cite{Alwall:2014hca},
with initial state radiation modeled 
using the {\tt MGISR} plugin~\cite{Li:2018qnh}.
The background cross-sections are calculated 
with the respective generators,
except for {\tt KKMC}, 
for which results from~\cite{Banerjee:2007is} are used.
The detector simulations and reconstructions are performed with 
{\tt GEANT3}~\cite{Brun:1987ma} and 
{\tt BASF}~\cite{Itoh:1997st}, respectively.


An important aspect of this analysis,
in which it differs from the previous search performed 
by the \babar experiment~\cite{BaBar:2020jma},
is background modeling using MC samples and data in control regions.
We use the multivariate analysis technique 
to enhance the presence of
the signal over the background, as well as to define control regions,
corresponding to regions enriched with each background component.
The normalizations of the backgrounds are obtained
by fitting the different MC components to data in 
different control regions.
Studies of \epem and \mumu invariant masses 
as the discriminating variables
are carried out by blinding the signal region
until the optimization of the selection criteria is complete.
In the final set of fits in the signal region,
a uniform shape with Poisson fluctuations 
is added as an additional component
to account for background from the unsimulated SM four-lepton processes
$\epem \to \tautau\epem$ and $\epem \to \tautau\mumu$.


We look for events with four tracks,
each selected with a systematic uncertainty 
on the tracking efficiency
of 0.35\%~\cite{Belle:2012iwr}.
To suppress mis-reconstructed and beam-induced tracks,
we require the transverse (${dr}$) and longitudinal (${|dz|}$)
projection of the distances of the closest approach 
to the interaction point (IP) 
be smaller than 10~\mm and 50~\mm, respectively.
This selection criteria probes the parameter space 
with $\xi \sim 1$, which corresponds to a decay length of 
$\phi_L$ less than $\sim$ 10~\mm.
For the $\m_{\phi_L} < 0.1~\gev$ region, decay lengths can be larger than 10~\mm.
In such cases, we require looser criteria of: ${dr} < 50~\mm$, and ${|dz|} < 50~\mm$.

The net charge of the event is required to be zero.
In the $\phi_L \to \epem$ ($\mumu$) channel, 
we require at least one track to be identified as \ep (\mup)
and one track to be identified as \en (\mun)
by our particle identification (PID) system.
Correction factors for efficiency and the mis-identification rates 
are obtained using control samples from data, and applied to MC.
The precision of these correction factors is included 
as a systematic uncertainty.

We reconstruct $\phi_L$ candidates
by fitting each pair of \epem or \mumu
to a common vertex,
while the remaining tracks in the event
come from 1-prong decays of the two $\tau$ leptons.
Between $25\%$ and $50\%$ of the signal events  
have more than one $\phi_L$ candidate,
with the average multiplicity decreasing 
from $1.7$ to $1.3$ at higher $m_{\phi_L}$ values.
We choose the candidate with the smallest opening angle 
in the laboratory frame
to ensure there is exactly one $\phi_L$ per selected event.
The efficiency to select the true $\phi_L$ candidate per signal event
is more than 98\% (83\%) 
for $\phi_L \to \epem$ $(\mumu)$ channel.


The major background for $\phi_L \to \epem$ search comes 
from $\epem\to\tautau$ events,
where one of the $\tau^\pm$ leptons decays 
into a $\rho^\pm$ producing a $\pi^0$,
which decays into $e^+e^-\gamma$ final state,
thereby faking the event topology of the signal.
The major background for $\phi_L \to \mumu$ 
search till $m_{\phi_L}=1~\gev$
comes from $\epem\to\tautau$  events,
where one of the $\tau^\pm$ leptons decay contains 3 charged pions,
some of which are misidentified as muons. Beyond $m_{\phi_L}=1~\gev$,
the two muons mostly come from semileptonic decays of heavy quarks in $\epem\to\qqbar$ events.

To suppress most of the Bhabha, \mumu and two-photon backgrounds,
we use rectangular selection criteria on the 2-dimensional plane:
$M_{\mathrm{miss}} \in [2, 6]~\gev$ and 
$\theta_{\mathrm{miss}}^{\mathrm{CM}} \in [30^\circ, 150^\circ]$,
where the missing mass $(M_{\mathrm{miss}})$ is evaluated using  
the four tracks and all neutrals detected in the final state,
and $\theta_{\mathrm{miss}}^{\mathrm{CM}}$ 
is the polar angle of the missing momentum in the c.m. frame.

\begin{figure}[t]
\includegraphics[width=.47\textwidth]{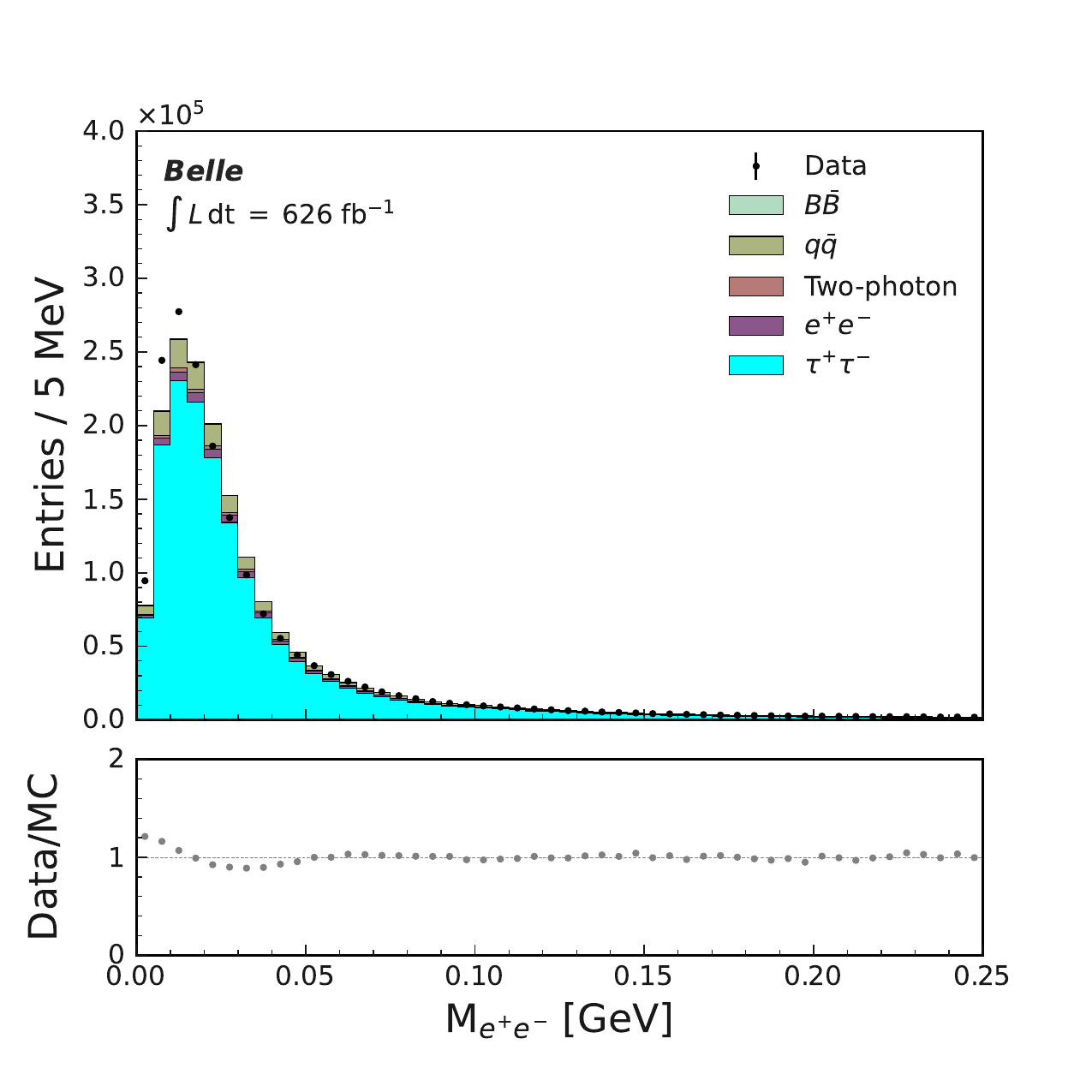}
\includegraphics[width=.47\textwidth]{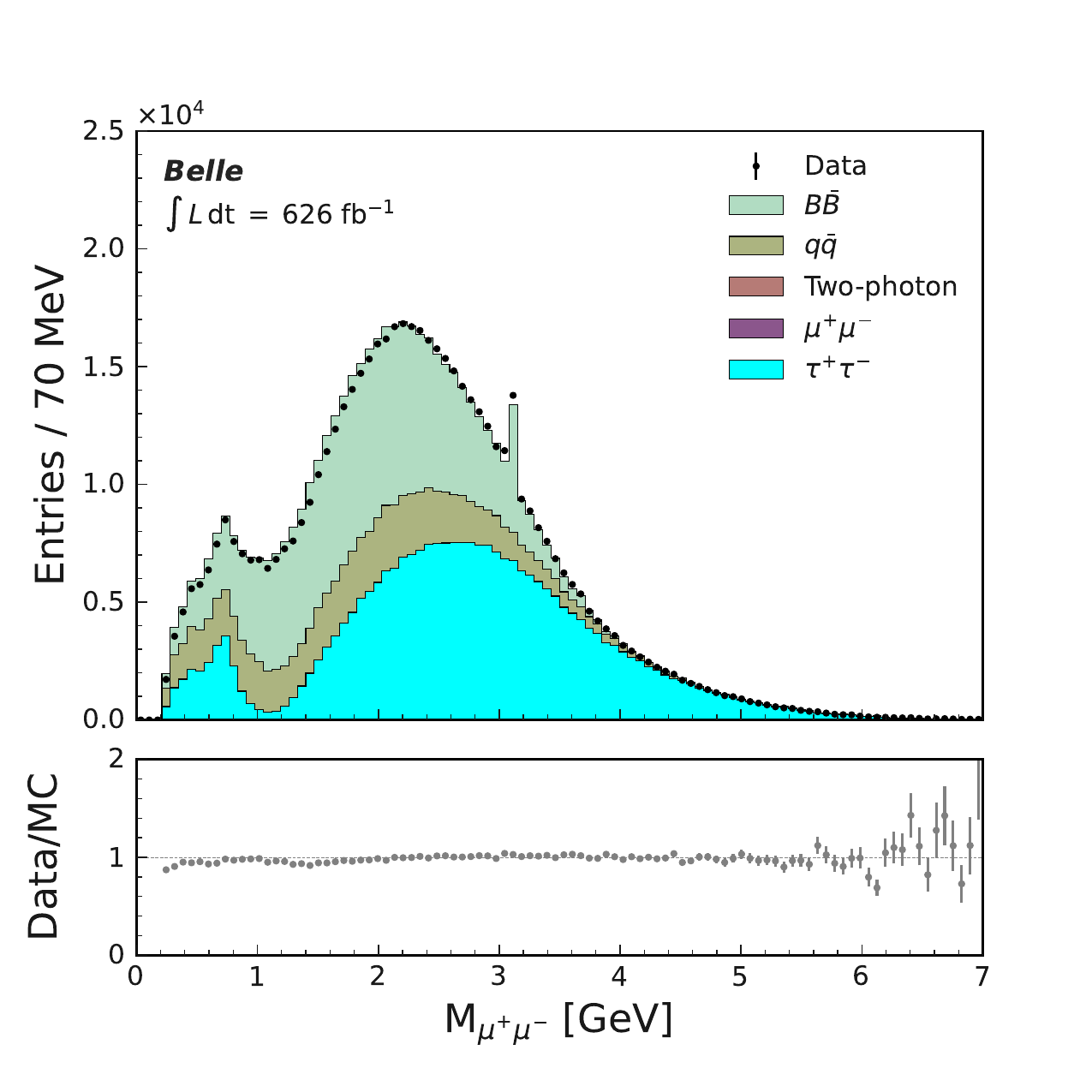}
\caption{Data and MC distributions of \epem invariant mass 
for $\phi_L \to \epem$ channel (top)
and \mumu invariant mass for $\phi_L \to \mumu$ channel (bottom) 
in the GCR.
All corrections and scale factors are applied to the 
MC distributions,
after normalizing them to the integrated luminosity of data.
}
\label{fig_data_mc_agreement_hl_InvM_CR}
\end{figure}

To suppress the remaining backgrounds, we train a multiclass boosted decision tree 
(BDT) for each channel,
using the \texttt{GradientBoostingClassifier} model available in 
\texttt{scikit-learn}~\cite{scikit-learn}.
We define five BDT scores to discriminate between the signal 
and four different types of backgrounds:
\tautau, \epem (\mumu), $q\bar{q}$ and $\BB$ in $\phi_L \to \epem$ ($\phi_L \to \mumu$) channel.

The top four variables ranked according to their feature importance 
in the BDT for the $\phi_L \to \epem$ channel are the thrust in the c.m. frame~\cite{Brandt:1964sa},
the opening angle between the daughters of $\phi_L$ candidate 
in the laboratory frame,
$M_{\mathrm{miss}}$ and the transverse distance of the vertex 
of the $\phi_L$ candidate from the IP.
The top four variables for the $\phi_L \to \mumu$ channel are
the invariant mass of \taup and \taum daughter tracks,
thrust, $M_{\mathrm{miss}}$ and the total energy of the 
reconstructed $\phi_L$ candidate in the laboratory frame.
The other variables used in the BDT are:
event shapes (ratios of Fox-Wolfram moments~\cite{Fox:1978vu}),
missing particles (visible energy,
and direction of missing momentum in the c.m. frame),
$\phi_L$ candidate (transverse momentum of daughter particles),
PID (number of leptons, pion-kaon discriminator 
for $\phi_L$ daughters),
neutral activity (number of $\pi^0$, 
the sum of energy deposited in ECL not associated with a track),
and invariant mass of the system formed 
by the $\phi_L$ candidate and its nearest photon.


In order to understand the background processes,
we define the general control region (GCR) 
with negligible signal contributions for each channel, 
by requiring the signal score to be less than 0.5.
The di-lepton mass distributions in the GCRs 
are shown in Fig.~\ref{fig_data_mc_agreement_hl_InvM_CR}. 
We obtain the scale factors for each background component
via a simultaneous fit across both channels.
In order to estimate the uncertainty of the scale factors, 
we define a special control region (SCR) for each of those four
backgrounds by requiring the corresponding BDT score 
to be greater than 0.5. 
We take the difference between the scale factors obtained from GCR 
and SCR as the uncertainty of each background contribution, 
except for the two-photon background, 
where the uncertainty is purely statistical.
For the dominant background processes of \tautau, 
the scale factor is consistent with unity, 
with 6\% (11\%) relative uncertainty in \epem (\mumu) channel.

\begin{figure}[t]
\vskip 0.3cm
\includegraphics[width=.47\textwidth]{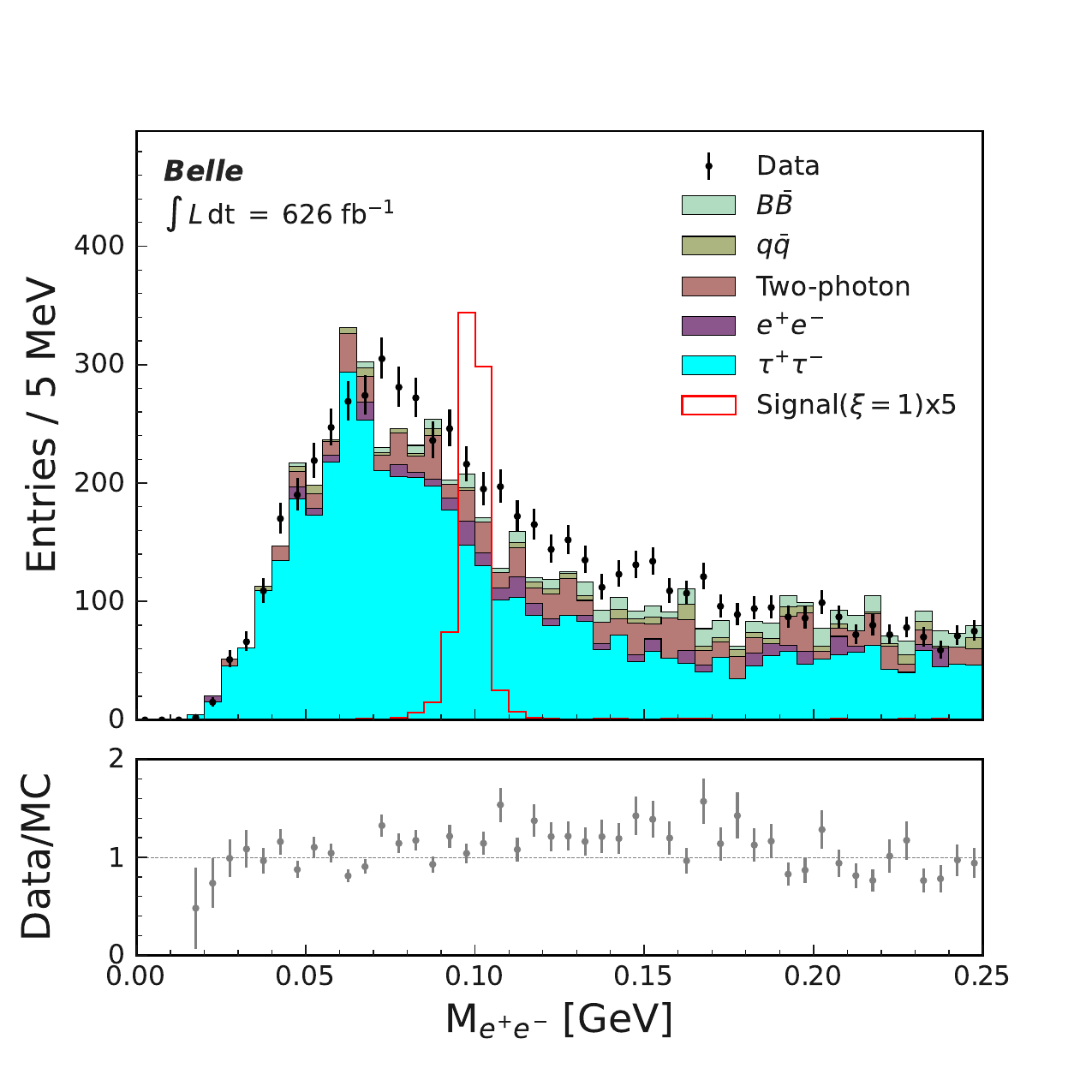}
\vskip 0.3cm
\includegraphics[width=.47\textwidth]{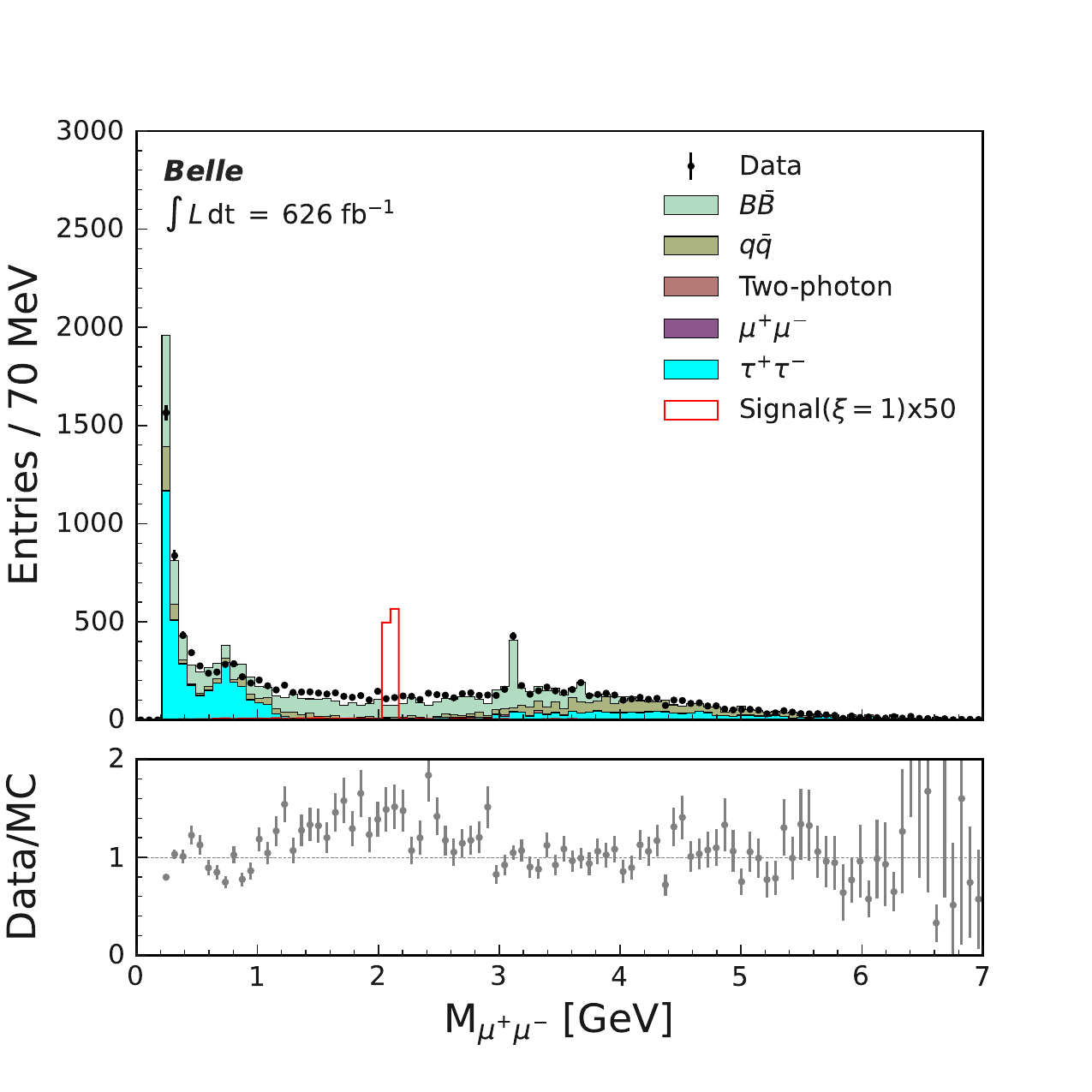}
\caption{Data and MC distributions of \epem invariant mass 
for $\phi_L \to \epem$ channel (top)
and \mumu invariant mass for $\phi_L \to \mumu$ channel (bottom) in SR.
The MC  are normalized to data, as in 
Fig.~\ref{fig_data_mc_agreement_hl_InvM_CR}.
The signal sample in $\phi_L \to \epem$ ($\phi_L \to \mumu$) channel is
generated with $m_{\phi_L}$ = 100~\mev (2.1~\gev).
}
\label{fig_data_mc_agreement_hl_InvM_SR}
\end{figure}

We define the signal region (SR)
with signal score $>$ 0.95 (0.65),
as an optimum choice that maximizes the sensitivity
for the \epem (\mumu) channels,
where the signal efficiency varies between 0.5\% to 7.5\% (5\% to 17\%).
The distributions of \epem and \mumu invariant mass  in SR
are shown in Fig.~\ref{fig_data_mc_agreement_hl_InvM_SR},
along with the MC backgrounds (stacked
histograms) and signal distributions (red histograms).
The ratio between the data and the sum of the MC backgrounds is 
shown at the bottom of each figure.
No obvious narrow peak structure is observed, 
except for the \jpsi signal in the \mumu channel.
A slight excess of data above the MC samples in both channels
is expected due to the above-mentioned unsimulated processes.

We search for narrow peaks in \epem (\mumu) invariant mass
distributions by performing binned maximum likelihood fits,
where the likelihood is defined as a product of Possion distributions
with expected events obtained from template histograms,
and Gaussian distributions describing systematic uncertainties,
as implemented in {\tt{HistFactory}}~\cite{Cranmer:2012sba}.
We use one bin from $2m_e$ ($2m_\mu$) to $m_{\phi_L} - 2 \sigma_{\phi_L}$,
2 to 8 bins in $m_{\phi_L} \pm 2 \sigma_{\phi_L}$ window,
and one bin from $m_{\phi_L} + 2 \sigma_{\phi_L}$ to 250~\mev (7~\gev).
Here $\sigma_{\phi_L}$ is the resolution of the \ellell mass
distribution for the signal, 
and it varies in the $[5,30]~\mev$ range,
increasing at larger values of $m_{\phi_L}$.
The mass of $\phi_L$ is kept fixed in the fit and scanned from
40~\mev to 210~\mev at 10~\mev intervals for \epem channel,
and from 225~\mev to 6.5~\gev at 25~\mev intervals for the \mumu 
channel. We skip the $\pm 50~\mev$ window around 
the nominal mass of \jpsi and \psitwos,
where we expect peaking backgrounds.
The fit includes systematic uncertainties 
from luminosity, tracking efficiency, 
momentum scale and PID corrections of $\phi_L$ daughter tracks,
scale factors and selection efficiency of BDT.
To account for the unsimulated processes,
we include an additional uniform background component.

\begin{figure}[!htbp]
\includegraphics[width=.47\textwidth]{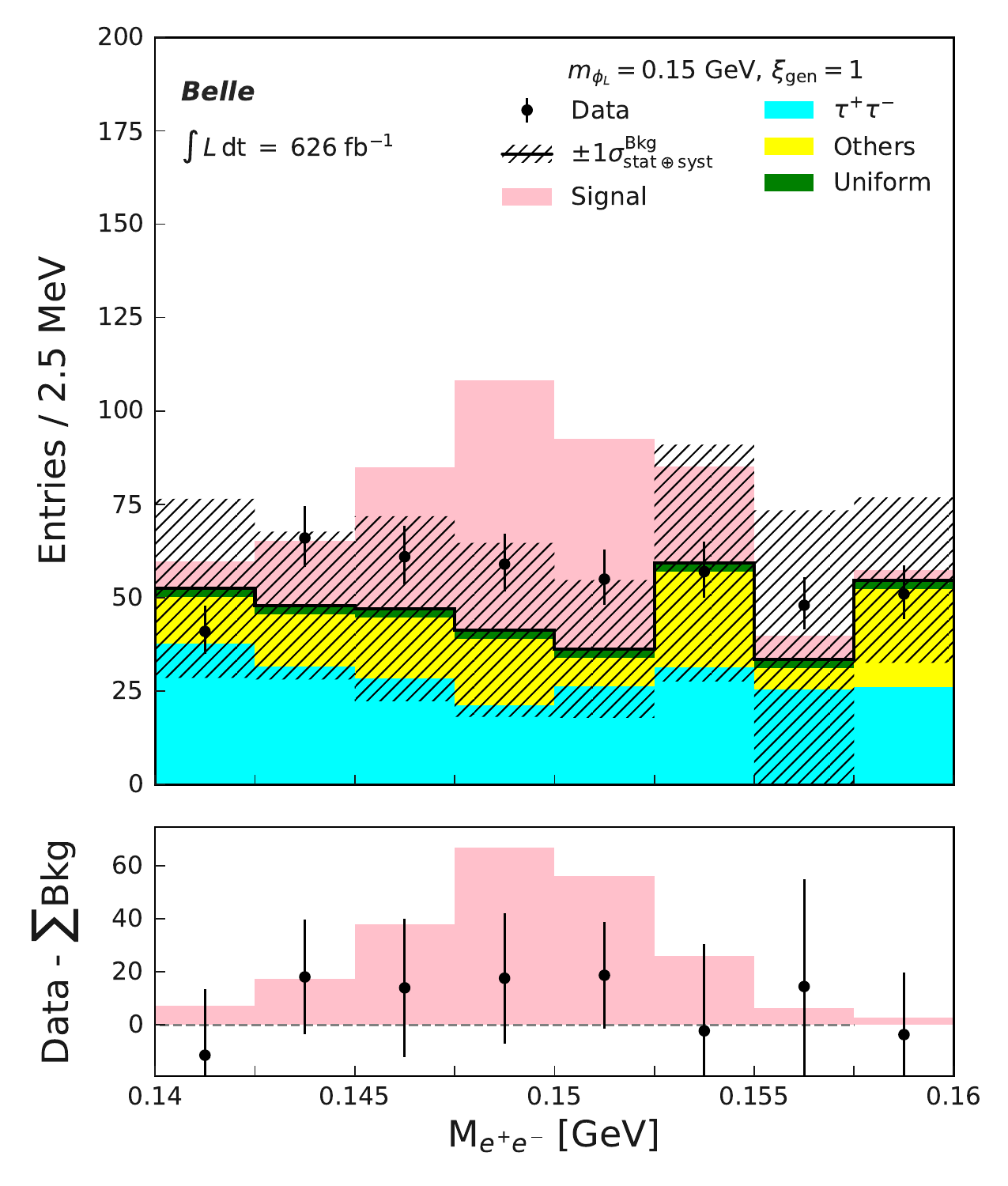}
\caption{$e^+ e^-$ invariant mass distributions are shown in the top sub-plot
with 2.5~\mev bin-width in the signal region corresponding to $m_{\phi_L}$ = 150~\mev,
which has the highest observed significance of 2.5 standard deviations in this channel.
The data are shown as black dots, while the signal, $\tautau$, other Monte Carlo
components of the backgrounds and the additional uniform background component
are shown by pink, cyan, yellow and green histograms, respectively.
The statistical and systematic components of the uncertainty on total background
have been added in quadrature and are shown by the shaded histogram.
The bottom sub-plot compares the signal distribution with data minus the background contributions.
}
\label{hl_ws_ee}
\end{figure}
\begin{figure}[!htbp]
\includegraphics[width=.47\textwidth]{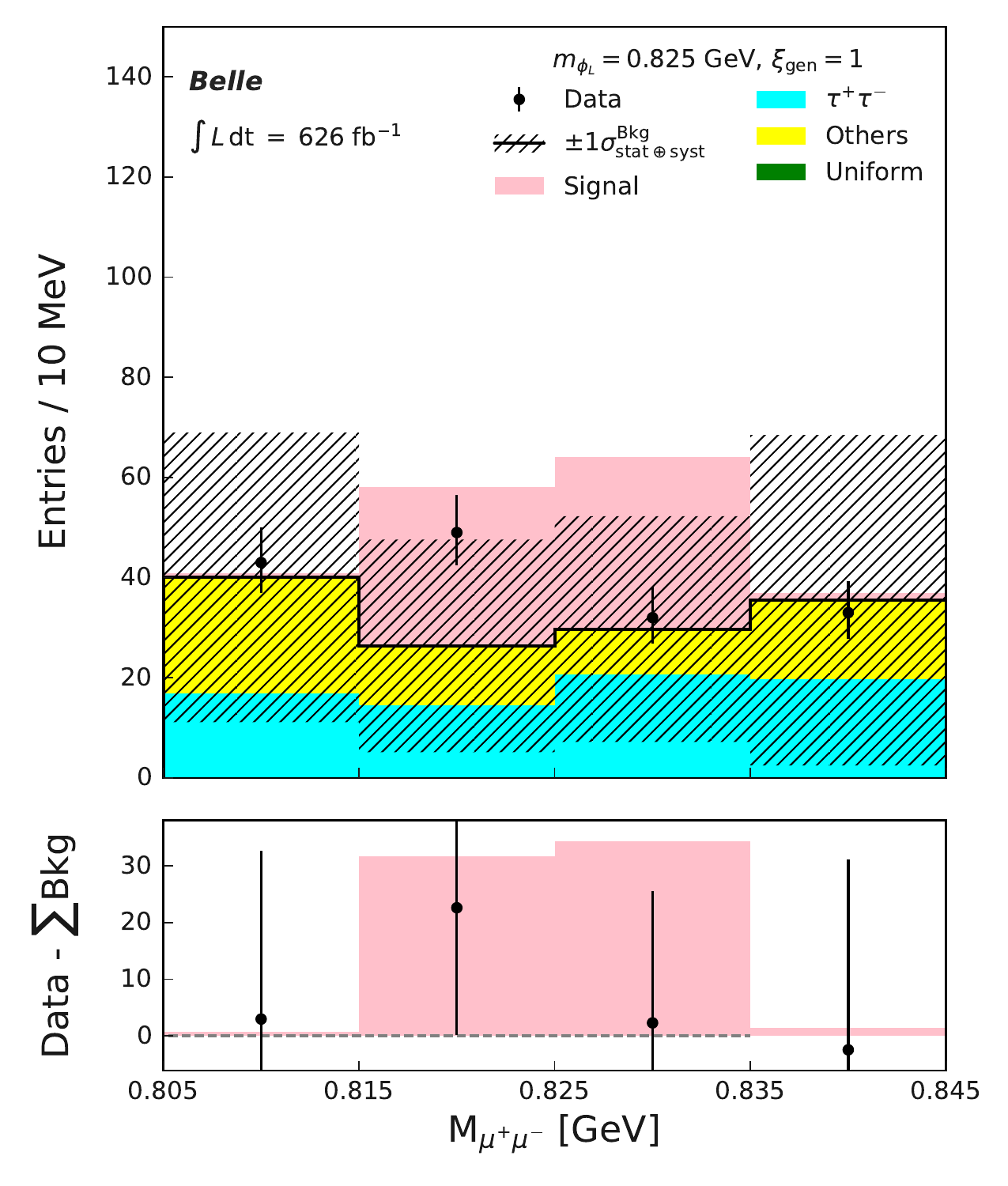}
\includegraphics[width=.47\textwidth]{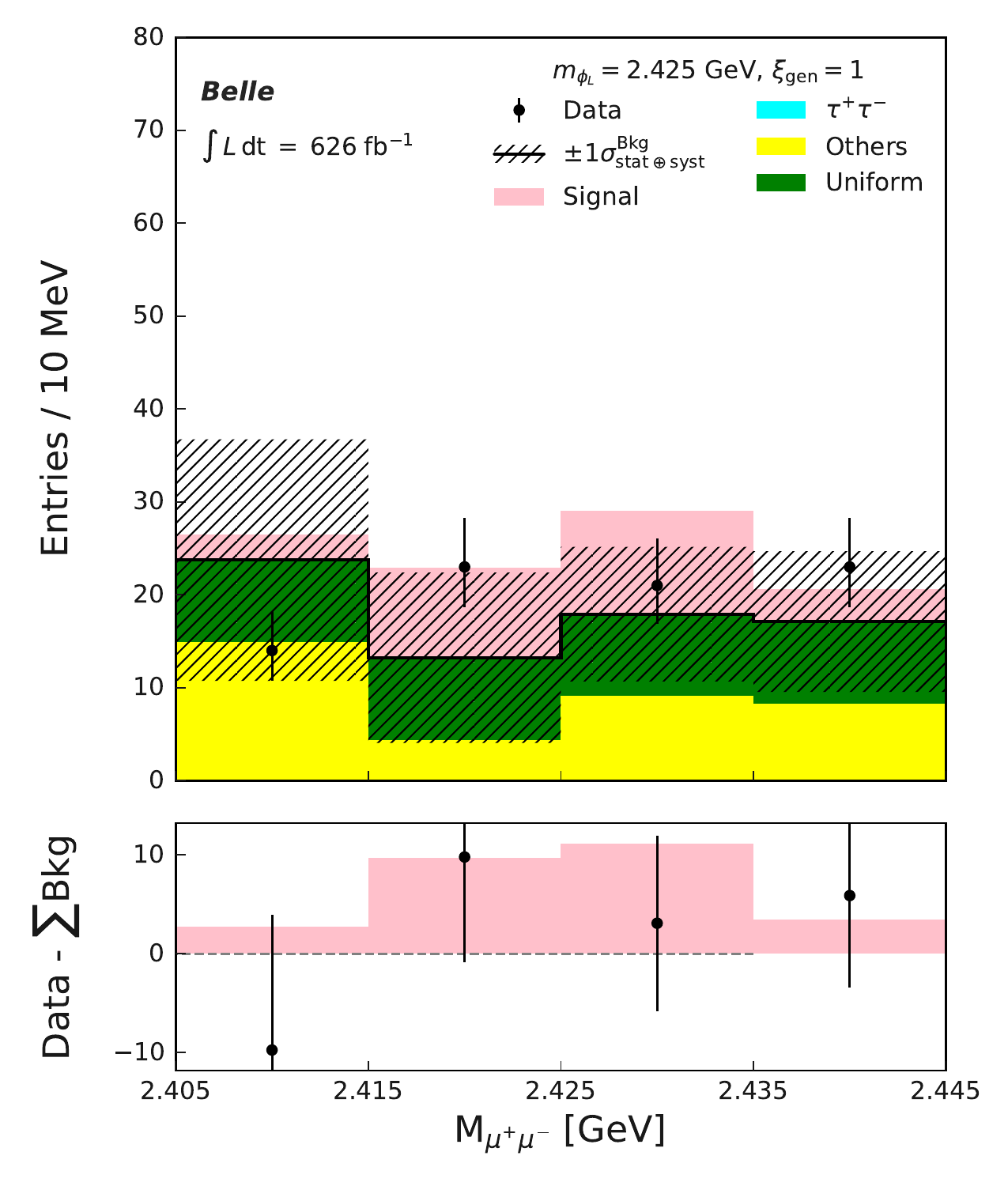}
\caption{
$\mu^+\mu^-$ invariant mass distributions are shown in the top inset
with 10~\mev bin-width in the signal region corresponding to $m_{\phi_L}$ = 0.825~\gev and 2.425~\gev,
which have the second highest and highest observed significance of 2.1 and 2.2 standard deviations in this channel, respectively.
The data are shown as black dots, while the signal, $\tautau$, other Monte Carlo
components of the backgrounds and the additional uniform background component
are shown by pink, cyan, yellow and green histograms, respectively.
The statistical and systematic components of the uncertainty on total background
have been added in quadrature and are shown by the shaded histogram.
The bottom sub-plots compare the signal distribution with data minus the background contributions.
}
\label{hl_ws_mumu}
\end{figure}
\begin{figure}[!htbp]
\includegraphics[width=.47\textwidth]{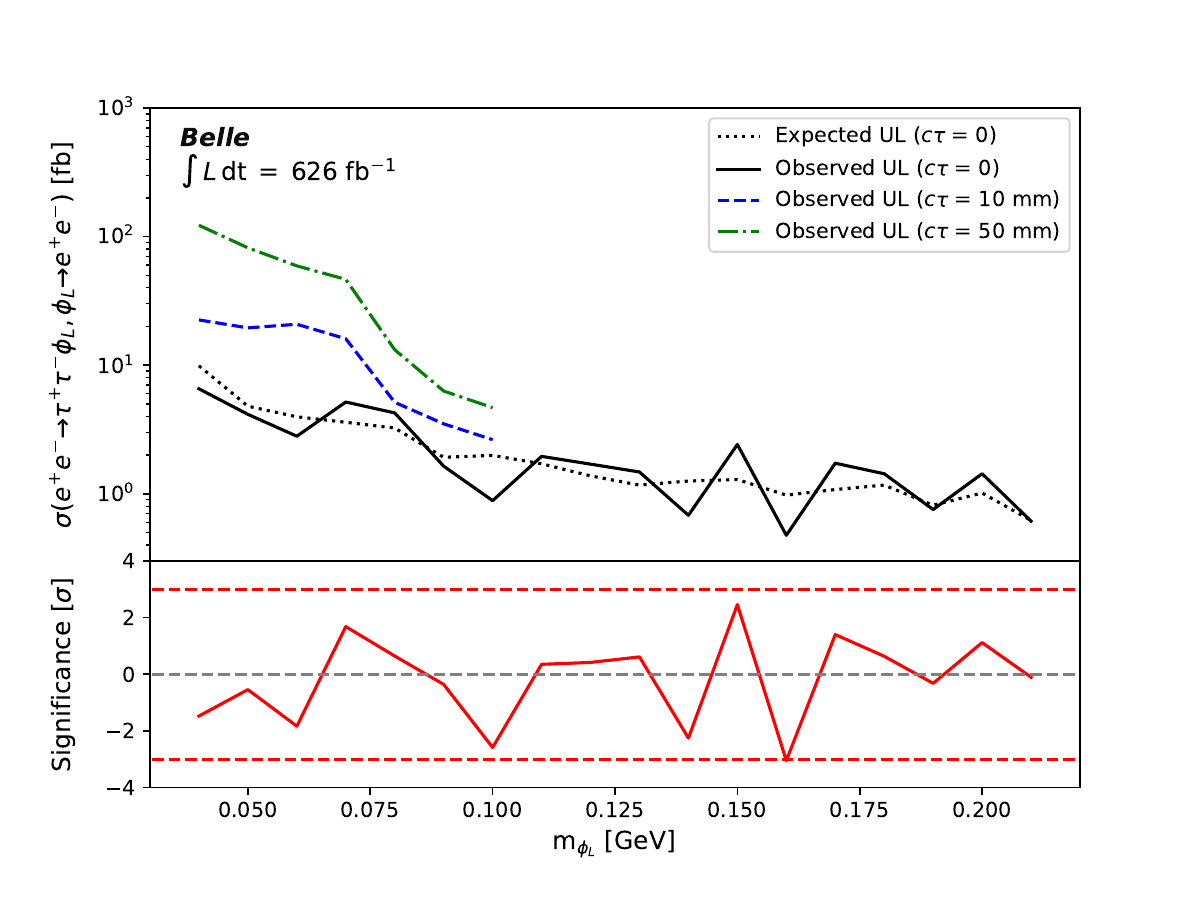}
\includegraphics[width=.47\textwidth]{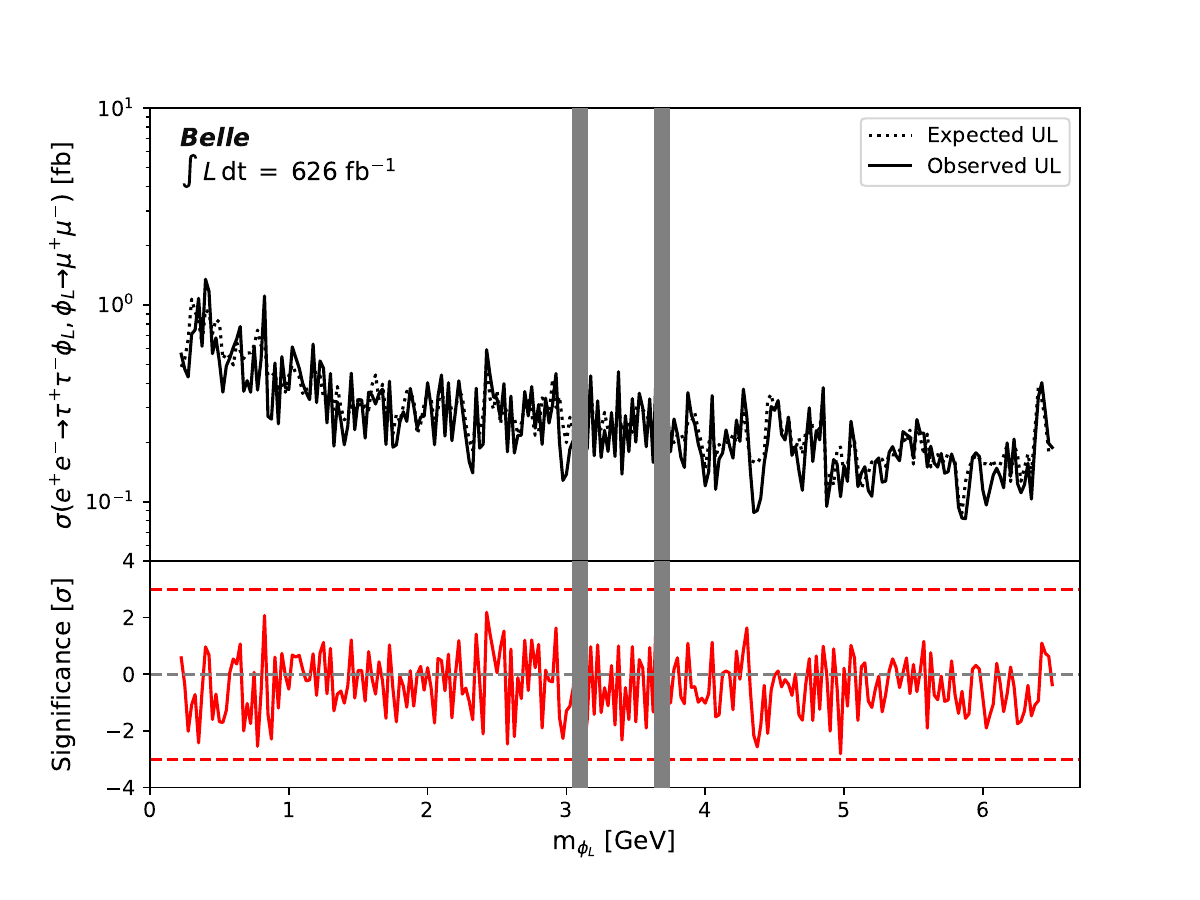}
\caption{
  Observed upper limits at 90\% CL on the signal cross-section
    with mean proper decay lengths $(c\tau)$ of 0~\mm, 10~\mm and 50~\mm, respectively, 
    are shown in the top sub-plots as a function of the dark leptophilic scalar mass
    for $\phi_L \to \epem$ channel (top) and $\phi_L \to \mumu$ channel (bottom).
    The bottom sub-plots in both of the figures show the observed significance for each channel.
    See text for details.}
\label{fig_cross_UL_with_obs_sig}
\end{figure}

We use the profile likelihood ratio as the test
statistic~\cite{Cowan:2010js}
to compare data with signal-plus-background 
or background-only hypothesis.
The fraction of each background component
and the additional uniform component
are allowed to vary within their uncertainties.
The fit returns the signal yield as well as 
the normalization factor for each background component,
along with the nuisance parameters 
describing systematic uncertainties.
In order to obtain the signal significance, we first calculate the 
p-value, the probability that the data is explained as 
the statistical fluctuation of the background.
We then calculate the signed significance,
where the sign follows the convention elaborated in Ref.~\cite{ATLAS:2012qaq}.
The signal significances are shown in the bottom sub-plots 
in Fig.~\ref{fig_cross_UL_with_obs_sig} for $\epem$ (top)
and $\mumu$ (bottom) channels.
We find all the mass points have significance less than 
3 standard deviations ($\sigma$).
Fit results for 3 mass points ($m_{\phi_L}$ = 
0.15~\gev, 0.825~\gev and 2.425~\gev)
with significance more than $2~\sigma$ are shown 
in Fig.~\ref{hl_ws_ee} and Fig.~\ref{hl_ws_mumu}.

To enable direct comparison with existing upper limits (UL) from the \babar experiment~\cite{BaBar:2020jma},
we calculate Bayesian UL using the Metropolis-Hastings algorithm~\cite{Metropolis:1953am,Hastings:1970} as implemented in RooStats~\cite{Moneta:2010pm}.
A toy Monte Carlo based numerical integration technique~\cite{Moneta:2010pm} is used to cross-check our Bayesian result, which agrees within a couple of percents across the whole mass range.
The UL of the signal cross-section at 90\% confidence level (CL)~\cite{Bayesian-footnote} are shown in the top sub-plots in 
Fig.~\ref{fig_cross_UL_with_obs_sig}.

We also cross-check our results using an alternate fitting method as used by \babar experiment~\cite{BaBar:2020jma},
where the background is modeled as a smooth polynomial from sideband data, and the signal is modeled as a Gaussian.
The observed significance from these two methods agree within $0.35~\sigma$ on the average.

Both the cross-section and the proper decay length ($c\tau$) of the dark leptophilic scalar depend on the coupling constant $\xi$.
For $m_{\phi_L}$ $>$ 0.1~\gev, the obtained UL on $\xi$ 
is consistent with the assumption that $c\tau$ is short 
enough to have negligible influence on the
signal detection efficiency.
However, for $m_{\phi_L}$ $<$ 0.1~\gev, $c\tau$ of $\sim$ 
10 to 50~\mm is expected for $\xi\sim 1$.
To take this dependence into account, we simulate the events 
with two additional values of $c\tau$ $=$ 10~\mm 
and $c\tau$ $=$ 50~\mm,
and re-perform the entire analysis to determine 
the UL on the cross-section for these values.
Using these UL and the known relation between $c\tau$ 
and $\xi$, we iteratively determine the UL on the $\xi$,
as shown in the top sub-plot of Fig.~\ref{fig_cross_UL_with_obs_sig}.

\begin{figure}[!htbp]
\includegraphics[width=.47\textwidth]{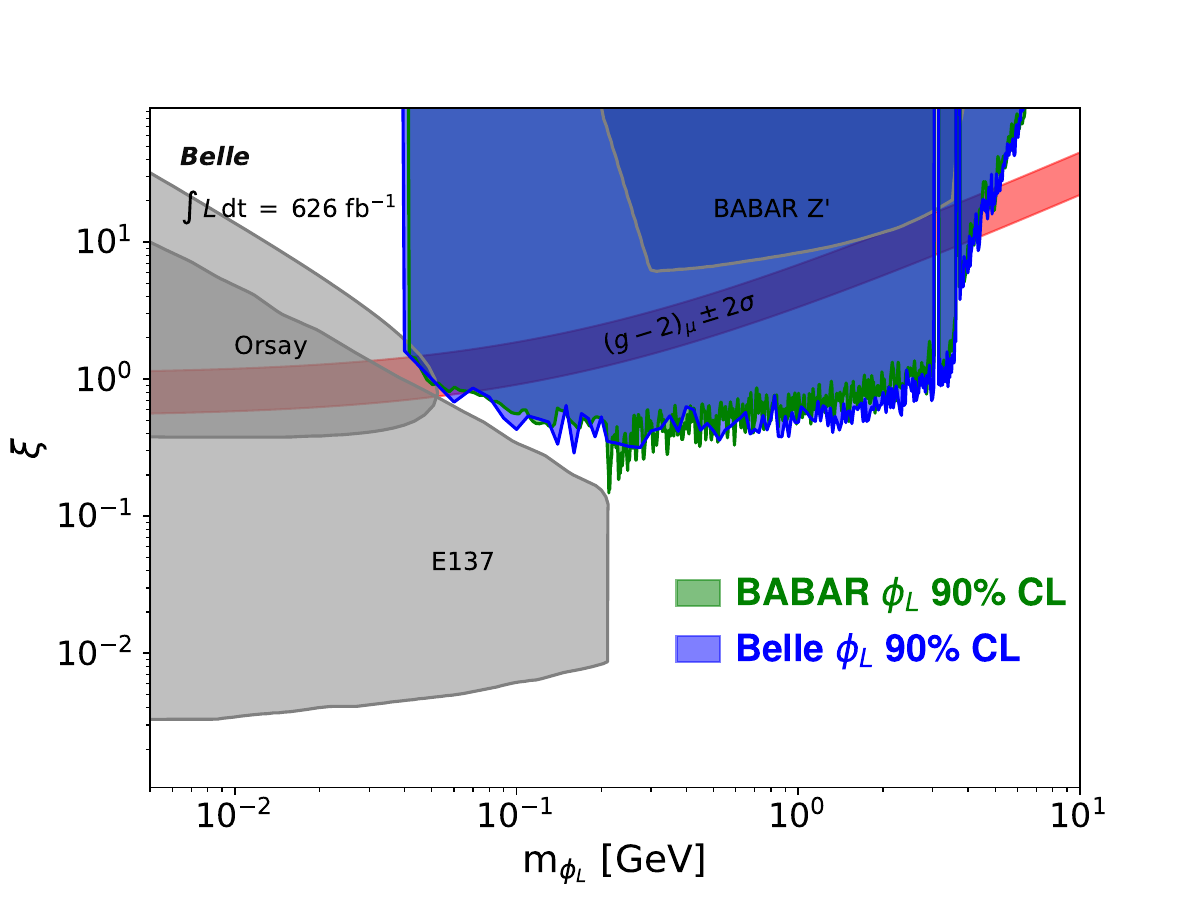} 
\caption{
  Observed upper limits at 90\% CL on the coupling constant $\xi$ 
  as a function of the $\phi_L$ mass from our search (blue), 
  overlaid with results from \babar (green)~\cite{BaBar:2020jma} 
  and other searches (gray)~\cite{BaBar:2016sci,Bjorken:1988as,Davier:1989wz}.
  The parameter space preferred by the $(g-2)_\mu$ measurement~\cite{Muong-2:2021ojo}
  is shown as a red band.}
\label{fig_Xi_UL_vs_m_phi}
\end{figure}

The exclusion region of the coupling constant $\xi$ vs. $m_{\phi_L}$
is shown in Fig.~\ref{fig_Xi_UL_vs_m_phi}, 
overlaid with previous results~\cite{BaBar:2016sci,Bjorken:1988as,Davier:1989wz,BaBar:2020jma}.
Our limits are tabulated in Table~\ref{UL_table_num}.

\begin{table*}[!htbp]
\begin{center}
\begin{tabular}{|cr|cr|cr|cr|cr|}
\hline
 $\mathrm{m}_{\phi_L}$ [GeV] &  $\xi^{\mathrm{UL}}_{\mathrm{obs}}$ &  $\mathrm{m}_{\phi_L}$ [GeV] &  $\xi^{\mathrm{UL}}_{\mathrm{obs}}$ &  $\mathrm{m}_{\phi_L}$ [GeV] &  $\xi^{\mathrm{UL}}_{\mathrm{obs}}$ &  $\mathrm{m}_{\phi_L}$ [GeV] &  $\xi^{\mathrm{UL}}_{\mathrm{obs}}$ &  $\mathrm{m}_{\phi_L}$ [GeV] &  $\xi^{\mathrm{UL}}_{\mathrm{obs}}$ \\
 \hline \hline
0.040 &     1.610 & 1.100 &     0.528 & 2.425 &     1.169 & 3.950 &     6.882 & 5.275 &    27.814 \\
0.050 &     1.000 & 1.125 &     0.506 & 2.450 &     1.017 & 3.975 &     6.742 & 5.300 &    28.678 \\
0.060 &     0.680 & 1.150 &     0.492 & 2.475 &     0.909 & 4.000 &     5.958 & 5.325 &    34.832 \\
0.070 &     0.860 & 1.175 &     0.690 & 2.500 &     0.907 & 4.025 &     6.761 & 5.350 &    36.915 \\
0.080 &     0.740 & 1.200 &     0.495 & 2.525 &     0.817 & 4.050 &    11.036 & 5.375 &    35.949 \\
0.090 &     0.520 & 1.225 &     0.640 & 2.550 &     1.014 & 4.075 &     6.652 & 5.400 &    35.807 \\
0.100 &     0.430 & 1.250 &     0.621 & 2.575 &     0.687 & 4.100 &     8.289 & 5.425 &    43.741 \\
0.110 &     0.539 & 1.275 &     0.456 & 2.600 &     0.963 & 4.125 &     8.929 & 5.450 &    44.129 \\
0.120 &     0.512 & 1.300 &     0.616 & 2.625 &     0.698 & 4.150 &    10.629 & 5.475 &    44.814 \\
0.130 &     0.486 & 1.325 &     0.408 & 2.650 &     0.782 & 4.175 &    10.067 & 5.500 &    40.440 \\
0.140 &     0.335 & 1.350 &     0.534 & 2.675 &     0.791 & 4.200 &     9.697 & 5.525 &    52.190 \\
0.150 &     0.642 & 1.375 &     0.485 & 2.700 &     1.033 & 4.225 &    12.562 & 5.550 &    49.473 \\
0.160 &     0.289 & 1.400 &     0.425 & 2.725 &     0.907 & 4.250 &    11.485 & 5.575 &    50.988 \\
0.170 &     0.560 & 1.425 &     0.481 & 2.750 &     1.088 & 4.275 &    16.106 & 5.600 &    42.933 \\
0.180 &     0.516 & 1.450 &     0.663 & 2.775 &     0.863 & 4.300 &    14.171 & 5.625 &    49.827 \\
0.190 &     0.380 & 1.475 &     0.482 & 2.800 &     1.003 & 4.325 &    10.710 & 5.650 &    46.459 \\
0.200 &     0.530 & 1.500 &     0.581 & 2.825 &     0.802 & 4.350 &     8.647 & 5.675 &    46.820 \\
0.210 &     0.351 & 1.525 &     0.586 & 2.850 &     1.060 & 4.375 &     9.011 & 5.700 &    51.991 \\
0.225 &     0.343 & 1.550 &     0.474 & 2.875 &     0.930 & 4.400 &     9.983 & 5.725 &    47.706 \\
0.250 &     0.323 & 1.575 &     0.627 & 2.900 &     1.042 & 4.425 &    12.643 & 5.750 &    49.518 \\
0.275 &     0.317 & 1.600 &     0.623 & 2.925 &     1.268 & 4.450 &    14.789 & 5.775 &    56.750 \\
0.300 &     0.418 & 1.625 &     0.599 & 2.950 &     0.860 & 4.475 &    18.627 & 5.800 &    55.059 \\
0.325 &     0.438 & 1.650 &     0.644 & 2.975 &     0.697 & 4.500 &    18.769 & 5.825 &    43.879 \\
0.350 &     0.539 & 1.675 &     0.669 & 3.000 &     0.728 & 4.525 &    20.492 & 5.850 &    42.688 \\
0.375 &     0.416 & 1.700 &     0.489 & 3.025 &     0.858 & 4.550 &    17.201 & 5.875 &    44.020 \\
0.400 &     0.629 & 1.725 &     0.717 & 3.150 &     0.912 & 4.575 &    17.250 & 5.900 &    53.655 \\
0.425 &     0.599 & 1.750 &     0.492 & 3.175 &     1.412 & 4.600 &    20.211 & 5.925 &    67.005 \\
0.450 &     0.424 & 1.775 &     0.504 & 3.200 &     0.892 & 4.625 &    16.634 & 5.950 &    71.073 \\
0.475 &     0.475 & 1.800 &     0.584 & 3.225 &     1.243 & 4.650 &    18.014 & 5.975 &    71.839 \\
0.500 &     0.420 & 1.825 &     0.626 & 3.250 &     0.904 & 4.675 &    16.081 & 6.000 &    61.226 \\
0.525 &     0.359 & 1.850 &     0.597 & 3.275 &     1.072 & 4.700 &    14.723 & 6.025 &    58.011 \\
0.550 &     0.425 & 1.875 &     0.733 & 3.300 &     0.959 & 4.725 &    19.830 & 6.050 &    65.471 \\
0.575 &     0.453 & 1.900 &     0.667 & 3.325 &     1.221 & 4.750 &    25.134 & 6.075 &    74.199 \\
0.600 &     0.490 & 1.925 &     0.595 & 3.350 &     0.953 & 4.775 &    18.791 & 6.100 &    79.979 \\
0.625 &     0.524 & 1.950 &     0.641 & 3.375 &     1.585 & 4.800 &    23.124 & 6.125 &    79.262 \\
0.650 &     0.572 & 1.975 &     0.654 & 3.400 &     0.880 & 4.825 &    22.510 & 6.150 &    77.155 \\
0.675 &     0.398 & 2.000 &     0.802 & 3.425 &     1.257 & 4.850 &    31.242 & 6.175 &   104.137 \\
0.700 &     0.430 & 2.025 &     0.706 & 3.450 &     1.029 & 4.875 &    16.014 & 6.200 &    88.988 \\
0.725 &     0.409 & 2.050 &     0.570 & 3.475 &     1.422 & 4.900 &    18.705 & 6.225 &   115.162 \\
0.750 &     0.543 & 2.075 &     0.767 & 3.500 &     1.116 & 4.925 &    22.153 & 6.250 &    92.529 \\
0.775 &     0.426 & 2.100 &     0.876 & 3.525 &     1.504 & 4.950 &    22.205 & 6.275 &    91.434 \\
0.800 &     0.516 & 2.125 &     0.618 & 3.550 &     1.394 & 4.975 &    18.850 & 6.300 &   100.179 \\
0.825 &     0.759 & 2.150 &     0.855 & 3.575 &     1.314 & 5.000 &    23.159 & 6.325 &   118.988 \\
0.850 &     0.382 & 2.175 &     0.616 & 3.600 &     2.211 & 5.025 &    21.628 & 6.350 &   100.772 \\
0.875 &     0.380 & 2.200 &     0.735 & 3.625 &     1.877 & 5.050 &    31.480 & 6.375 &   146.809 \\
0.900 &     0.536 & 2.225 &     0.893 & 3.750 &     3.806 & 5.075 &    28.369 & 6.400 &   202.637 \\
0.925 &     0.380 & 2.250 &     0.765 & 3.775 &     5.047 & 5.100 &    22.649 & 6.425 &   231.121 \\
0.950 &     0.571 & 2.275 &     0.659 & 3.800 &     4.989 & 5.125 &    25.092 & 6.450 &   205.183 \\
0.975 &     0.483 & 2.300 &     0.573 & 3.825 &     4.700 & 5.150 &    26.815 & 6.475 &   180.166 \\
1.000 &     0.483 & 2.325 &     0.544 & 3.850 &     4.740 & 5.175 &    23.824 & 6.500 &   185.935 \\
1.025 &     0.628 & 2.350 &     0.902 & 3.875 &     7.846 & 5.200 &    23.664 &       &           \\
1.050 &     0.598 & 2.375 &     0.643 & 3.900 &     7.309 & 5.225 &    29.744 &       &           \\
1.075 &     0.569 & 2.400 &     0.666 & 3.925 &     7.322 & 5.250 &    31.265 &       &           \\
\hline
\end{tabular}

\caption{Observed upper limits at 90\% CL on the coupling constant $\xi$ as a function of the dark scalar mass.}
\label{UL_table_num}
\end{center}
\end{table*}

A fit to the ratio of limits obtained by the \babar experiment~\cite{BaBar:2020jma}
and our limits show that our results are more constraining by 19\% on the average.
We exclude the parameter space with $m_{\phi_L}$ 
between [0.04,4]~\gev favored by $(g-2)_\mu$ at 90\% 
CL~\cite{Batell:2016ove,Liu:2020qgx}.

In conclusion, we search for a dark leptophilic scalar 
and set the UL on the 
cross-section of $\epem \to \tautau\phi_L$, $\phi_L \to \epem$ process 
in the range $[0.6, 7]~\fb$ and
on the cross-section of $\epem \to \tautau\phi_L,~\phi_L \to \mumu$ process 
in the range $[0.1, 2]~\fb$ at 90\% CL.
There is no such leptophilic scalar 
with mass less than 4~\gev
that can explain the observed excess in $(g-2)_\mu$.

This work, based on data collected using the Belle detector, which was
operated until June 2010, was supported by 
the Ministry of Education, Culture, Sports, Science, and
Technology (MEXT) of Japan, the Japan Society for the 
Promotion of Science (JSPS), and the Tau-Lepton Physics 
Research Center of Nagoya University; 
the Australian Research Council including grants
DP210101900, 
DP210102831, 
DE220100462, 
LE210100098, 
LE230100085; 
Austrian Federal Ministry of Education, Science and Research (FWF) and
FWF Austrian Science Fund No.~P~31361-N36;
National Key R\&D Program of China under Contract No.~2022YFA1601903,
National Natural Science Foundation of China and research grants
No.~11575017,
No.~11761141009, 
No.~11705209, 
No.~11975076, 
No.~12135005, 
No.~12150004, 
No.~12161141008, 
and
No.~12175041, 
and Shandong Provincial Natural Science Foundation Project ZR2022JQ02;
the Ministry of Education, Youth and Sports of the Czech
Republic under Contract No.~LTT17020;
the Czech Science Foundation Grant No. 22-18469S;
Horizon 2020 ERC Advanced Grant No.~884719 and ERC Starting Grant No.~947006 ``InterLeptons'' (European Union);
the Carl Zeiss Foundation, the Deutsche Forschungsgemeinschaft, the
Excellence Cluster Universe, and the VolkswagenStiftung;
the Department of Atomic Energy (Project Identification No. RTI 4002) and the Department of Science and Technology of India; 
the Istituto Nazionale di Fisica Nucleare of Italy; 
National Research Foundation (NRF) of Korea Grant
Nos.~2016R1\-D1A1B\-02012900, 2018R1\-A2B\-3003643,
2018R1\-A6A1A\-06024970, RS\-2022\-00197659,
2019R1\-I1A3A\-01058933, 2021R1\-A6A1A\-03043957,
2021R1\-F1A\-1060423, 2021R1\-F1A\-1064008, 2022R1\-A2C\-1003993;
Radiation Science Research Institute, Foreign Large-size Research Facility Application Supporting project, the Global Science Experimental Data Hub Center of the Korea Institute of Science and Technology Information and KREONET/GLORIAD;
the Polish Ministry of Science and Higher Education and 
the National Science Center;
the Ministry of Science and Higher Education of the Russian Federation, Agreement 14.W03.31.0026, 
and the HSE University Basic Research Program, Moscow; 
University of Tabuk research grants
S-1440-0321, S-0256-1438, and S-0280-1439 (Saudi Arabia);
the Slovenian Research Agency Grant Nos. J1-9124 and P1-0135;
Ikerbasque, Basque Foundation for Science, Spain;
the Swiss National Science Foundation; 
the Ministry of Education and the Ministry of Science and Technology of Taiwan;
and the United States Department of Energy and the National Science Foundation.
These acknowledgements are not to be interpreted as an endorsement of any
statement made by any of our institutes, funding agencies, governments, or
their representatives.
We thank the KEKB group for the excellent operation of the
accelerator; the KEK cryogenics group for the efficient
operation of the solenoid; and the KEK computer group and the Pacific Northwest National
Laboratory (PNNL) Environmental Molecular Sciences Laboratory (EMSL)
computing group for strong computing support; and the National
Institute of Informatics, and Science Information NETwork 6 (SINET6) for
valuable network support.

\bibliography{Belle_Note_Journal_PRD_Final}

\begin{thebibliography}{45}
\bibitem [{Nat()}]{NaturalUnits}%
  \BibitemOpen
  \href@noop {} {}\bibinfo {organization} {Natural units ($\hbar$ = $c$ = 1)
  are used throughout this paper.}\BibitemShut {Stop}%
\end{thebibliography}
\bibliographystyle{apsrev4-1}

\end{document}